\documentclass[floatfix, twocolumn,pra,superscriptaddress]{revtex4-1}

\usepackage{graphicx}
\usepackage{amsmath}
\usepackage{subfigure}
\usepackage{dsfont}
\usepackage{hyperref}
\usepackage{braket}
\usepackage{siunitx}
\usepackage[T1]{fontenc}

\begin{document}

\title{Large spin shuttling oscillations enabling high-fidelity single qubit gates}

\author{Akshay Menon Pazhedath} 
\affiliation{Peter Grünberg Institute-Quantum Control (PGI-8), Forschungszentrum Jülich GmbH, D-52425 Jülich, Germany}
\affiliation{Institute for Theoretical Physics, University of Cologne, Zülpicher Straße 77, 50937 Cologne, Germany}
\author{Alessandro David}
\affiliation{Peter Grünberg Institute-Quantum Control (PGI-8), Forschungszentrum Jülich GmbH, D-52425 Jülich, Germany}
\author{Max Oberländer}
\affiliation{JARA-FIT Institute for Quantum Information, Forschungszentrum Jülich GmbH and RWTH Aachen University, Aachen, Germany}
\author{Matthias M. Müller} 
\affiliation{Peter Grünberg Institute-Quantum Control (PGI-8), Forschungszentrum Jülich GmbH, D-52425 Jülich, Germany}
\author{Tommaso Calarco}
\affiliation{Peter Grünberg Institute-Quantum Control (PGI-8), Forschungszentrum Jülich GmbH, D-52425 Jülich, Germany}
\affiliation{Institute for Theoretical Physics, University of Cologne, Zülpicher Straße 77, 50937 Cologne, Germany}
\affiliation{Dipartimento di Fisica e Astronomia, Università di Bologna, 40127 Bologna, Italy}
\author{Hendrik Bluhm} 
\affiliation{JARA-FIT Institute for Quantum Information, Forschungszentrum Jülich GmbH and RWTH Aachen University, Aachen, Germany}
\affiliation{ARQUE Systems GmbH, 52074 Aachen, Germany}
\author{Felix Motzoi}
\affiliation{Peter Grünberg Institute-Quantum Control (PGI-8), Forschungszentrum Jülich GmbH, D-52425 Jülich, Germany}

\begin{abstract}
  Semiconductor quantum dots have shown impressive breakthroughs in the last years, with single and two qubit gate fidelities matching other leading platforms and scalability still remaining a relative strength. However, due to qubit wiring considerations, mobile electron architectures have been proposed to facilitate upward scaling. In this work, we examine and demonstrate the possibility of significantly outperforming static EDSR-type single-qubit pulsing by taking advantage of the larger spatial mobility to achieve larger Rabi frequencies and reduce the effect of charge noise. Our theoretical results indicate that fidelities are ultimately bottlenecked by spin-valley physics, which can be suppressed through the use of quantum optimal control, and we demonstrate that, across different potential regimes and competing physical models, shuttling based single-qubit gates retain significant advantage over existing alternatives.

\end{abstract}

\maketitle

\section{Introduction}

Spin-based devices in semiconducting heterostructures \cite{burkardSemiconductorSpin2023} are promising candidates to host and manipulate qubits \cite{lossQuantumComputation1998a} as they offer advantages like inherent two-level structure, controllability through gate-tunable electrostatic potentials and direct interface with the semiconductor industry \cite{klemtElectricalManipulation2023, zwerverQubitsMade2022}. Si/SiGe heterostructures have been widely used to harbor spin qubits in quantum dots, with long spin relaxation time (beyond seconds)\cite{simmonsTunableSpin2011} and dephasing time (beyond tens of microseconds)\cite{kawakamiElectricalControl2014,neumannSimulationMicromagnet2015}. Moreover, semiconductor quantum dots have shown impressive breakthroughs in the last years, with single-qubit and two-qubit gate fidelities matching other leading platforms \cite{yonedaQuantumdotSpin2018,xueQuantumLogic2022, millsTwoqubitSilicon2022, noiriFastUniversal2022}.

Scalability of such systems is limited by the short range of the exchange interaction, which leads to a fan-out of electrostatic gates when wiring a large number of qubits, as well as to crosstalk errors. Sparse architectures with on-chip controls \cite{vandersypenInterfacingSpin2017, boterSpiderwebArray2022, kunneSpinBusArchitecture2023} are among the possible solutions, which require a long-range coherent link between distant qubit registers. We focus here on systems where the qubit is moved by a series of electrostatic gates allowing bi-directional conveyor-mode shuttling \cite{seidlerConveyormodeSingleelectron2021, langrockBlueprintScalable2023, xueSiSiGe2023, struckSpinEPRpairSeparation2023}. In this approach, the spin is manipulated by transporting it to a dedicated zone where a micro-magnet is present.

Single-qubit operations of quantum dot spins in presence of an artificial spin-orbit field have already been demonstrated \cite{nowackCoherentControl2007, pioro-ladriereElectricallyDriven2008, yonedaFastElectrical2014, kawakamiElectricalControl2014, yonedaQuantumdotSpin2018}. These experiments leverage electric-dipole spin resonance (EDSR) \cite{golovachElectricdipoleinducedSpin2006}, where the quantum dot confinement potential is oscillated beneath a static micro-magnet in order for the spin to experience a periodic magnetic field. However, this is achieved in a setting where the amplitude of oscillation is limited to $\sim$\SI{1}{\nano\metre}. This leads to larger gradient field strength requirements, which cause spin dephasing.

In contrast, shuttling architectures such as the SpinBus architecture \cite{kunneSpinBusArchitecture2023} is predicted to have the capability of performing larger amplitude oscillations, up to \num{10} or \SI{20}{\nano\metre}, weakening the gradient field requirement by at least an order of magnitude, and improving the spin dephasing time. This can be combined with a geometry-optimized micro-magnet for further improvements \cite{yonedaRobustMicromagnet2015, dumoulinstuyckLowDephasing2021}.

An obstacle to achieve high-fidelity gates in this context is represented by the presence of low lying valley states in the conduction band minimum of Si \cite{zwanenburgSiliconQuantum2013}. Evidence shows that these states couple to the spin degree of freedom (DOF) by having different g-factors \cite{ruskovElectronFactor2018,ferdousValleyDependent2018,veldhorstSpinorbitCoupling2015, rancicElectricDipole2016}. This modifies the standard fixed-frequency Rabi formula by introducing spatially dependent energy shifts, a fact that complicates the controllability and may ultimately lead to decoherence. The interaction between the electron wavefunction and the heterostructure walls lifts the degeneracy of the valley DOF, but the splitting is very sensitive to the local atomic arrangement \cite{wuetzAtomicFluctuations2022,friesenValleySplitting2007}. As the quantum dot is shuttled, the valley DOF may spend an undetected amount of time in its excited state, corresponding to a shift in the precession rate of the spin and to phase randomization.

In this work we model the shuttling-based EDSR operation, taking into account the dephasing mechanism between spin and valley states. A conveyor-mode shuttling device is modelled by generating position dependent valley eigenstates and eigenvalues, termed as valley landscape throughout this work, according to two distinct models. One model considers the presence of Ge atoms in the Si well (termed as the `Ge-diffusion model'), while the other focuses on the presence of fabrication miscuts along the length of the device (termed as the `step model')\cite{alessandro}. We show that the large electron oscillations increase the spin dephasing time of the qubit while operating at an order of magnitude lower magnetic gradient, with single-qubit gate fidelity limited by low valley splitting points $<$\SI{15}{\micro\electronvolt} (LVSPs) and large spin-valley coupling. The electron trajectory can be shaped using optimal control techniques presented in this work to improve the fidelity by at least two orders of magnitude, in a spin-valley physics agnostic manner. We study the dependence of the average gate fidelity on the driving frequency and the spatial amplitude of the trajectory to understand underlying physical effects, and we optimize over a parameter grid of pulse lengths and spatial amplitudes to show the advantage of optimal control in realizing fast high-fidelity gates. An extensive comparison of the valley models discussed above for various realistic scenarios, in terms of attainable fidelities, is also performed to elaborate on the advantage of using trajectory shaping over other methods. Our results indicate that shuttling-based EDSR is a viable and preferable way of performing single qubit operations in semiconducting platforms.

This paper is structured as follows. In Sec. \ref{Sec:Background}, we explain the relevant details of the SpinBus architecture and the model Hamiltonian used for our simulations, with a description of the different valley models considered. We follow this with a discussion on implementing the dynamics and optimization of the single-qubit operation, where we elucidate the optimal control methodology used. Sec. \ref{Sec:Results} deals with the main results of this work, where we show the advantages of a large-amplitude driven EDSR and compare the valley models based on the gate fidelities for a single prototypical device as well as for \num{1000} different sampled devices to ascertain the advantage of using optimal control in shaping electron trajectories. We discuss potential obstacles and future works in Sec. \ref{sec:Discussion} and conclude in Sec.~\ref{Sec:Conclusion}.
\begin{figure*}
    \centering
    \includegraphics[width=0.8\textwidth]{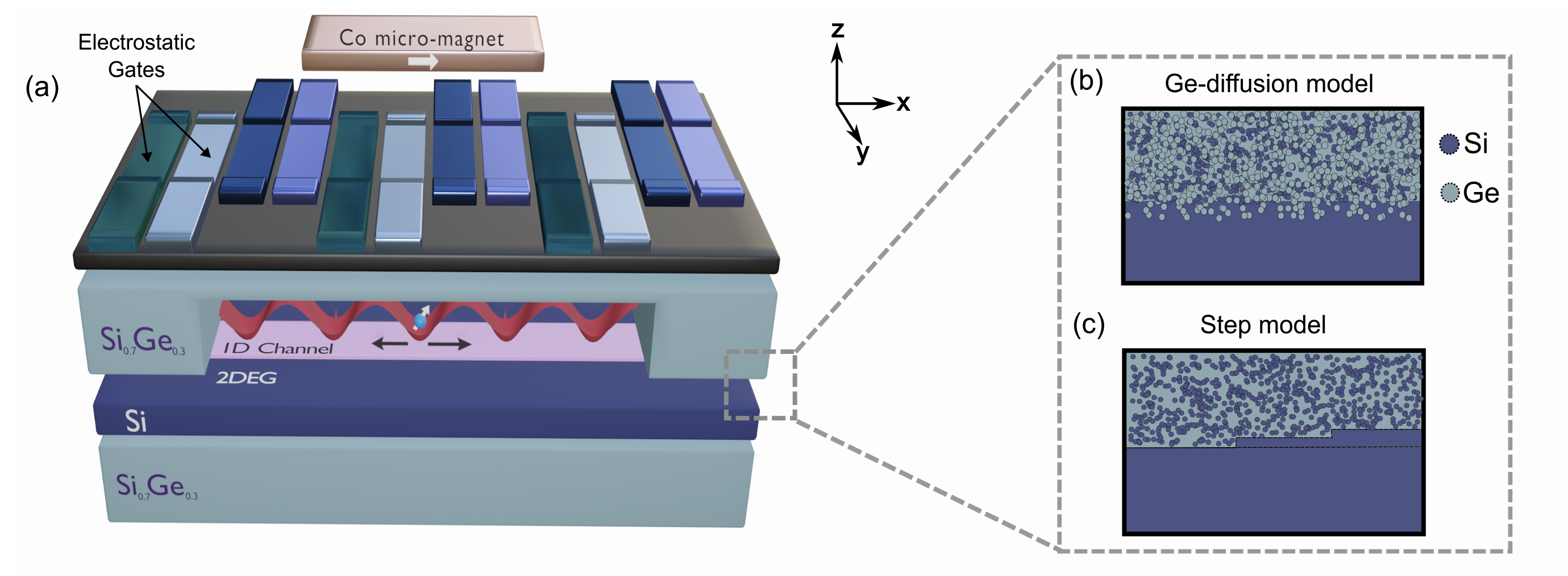}
    \caption{\textbf{Schematic of the manipulation zone in SpinBus.} (a) A suitable micro-magnet is placed on top of the electrostatic gates (same color gates receive the same phase shifted sinusoidal signal), which enables shuttling based single-qubit gates for the electron spin (blue ball) trapped in the Si quantum well. The trapping potential (shown in red) is moved back and forth to induce EDSR for the spin. The interface of SiGe and Si can be modelled in two different ways, as shown in (b) and (c). }
    \label{fig:manipulation-zone}
\end{figure*}

\section{Conveyor-mode single qubit rotations}\label{Sec:Background}
In this section, we model the manipulation zone of the quantum bus architecture (see Fig.~\ref{fig:manipulation-zone}) and we discuss the simulation and optimization of the single-qubit operation.

\subsection{Device architecture}
\label{sec:Device_architecture}
In a conveyor-mode spin-shuttling device \cite{langrockBlueprintScalable2023}, a row of electric gates generates an array of quantum dots with sinusoidal profile inside the quantum well of a Si/SiGe heterostructure. The transported qubit is encoded in the spin of a single electron trapped in one of these quantum dots. Bi-directional shuttling of the electron over $\sim$\SI{10}{\micro\metre} is possible \cite{xueSiSiGe2023} with only four phase shifted sine waves as control signals to the gates. Fig.~\ref{fig:manipulation-zone} illustrates such an apparatus. 

Single qubit gates are performed via EDSR at the so called manipulation zone
, by oscillating the electron under an external magnetic field ($\sim$\qtyrange[range-units = single, range-phrase=--]{20}{50}{\milli\tesla}) \cite{kunneSpinBusArchitecture2023} around a magnetic gradient generated by a suitable micro-magnet. In comparison, the high fidelity gates shown by Ref.~\cite{yonedaQuantumdotSpin2018} adopt a similar approach of using micro-magnets in conjunction with electron position oscillation to perform EDSR. The difference is in the choice of the operating regime, with a far higher external magnetic field ($\sim$\SI{0.5}{\tesla}), an order of magnitude higher gradient magnetic fields and an order of magnitude lower amplitude for oscillations.

\subsection{Model Hamiltonian}
\label{sec:Hamiltonian}

We build on the model presented in Ref.~\cite{alessandro} by adding an extra term describing the artificial spin-orbit field. The computational subspace of interest is the spin of the electron in the quantum dot, a two-level system (TLS). The spin can interact with the so called valley degree of freedom (DOF) in Si, also a TLS \cite{zwanenburgSiliconQuantum2013}. The combined Hilbert space can be written as
\begin{equation}
    \mathcal{H}=\mathcal{H}_{\text{valley}}\otimes\mathcal{H}_{\text{spin}},
\end{equation}
with $\ket{+k_z},\ket{-k_z}$ as the valley basis and ${\ket{0},\ket{1}}$ (computational basis) as the spin basis. Here, $\ket{+k_z},\ket{-k_z}$ correspond to the two low-lying valley states after the \num{6}-fold degeneracy in the conduction band of Si is lifted by strain.

The Hamiltonian has four terms,
\begin{equation}
    H = H_{\text{Zeeman}} + H_{\text{Rabi}} + H_{\text{valley}} + H_{\text{spin-valley}},
    \label{eq:Full Hamiltonian}
\end{equation}
where 
\begin{equation}
    H_{\text{Zeeman}} = \frac{1}{2}g\mu_{B} B_{z}~\mathds{1}\otimes\sigma_{z}
\end{equation}
with $g$ being the Landé g-factor of the electron ($\approx$\num{2}), $\mu_{B}$ being the Bohr magneton, $B_z$ being the constant external magnetic field applied across the manipulation zone and $\sigma_z$ is the $z$ Pauli matrix acting on the spin. The value of $B_z$ is chosen to be \SI{20}{\milli\tesla} in this work. The drive term,
\begin{equation}
    H_{\text{Rabi}} = \frac{1}{2}g\mu_{B} B_{x}(x_\text{qd}(t))~\mathds{1}\otimes\sigma_{x}
\end{equation}
corresponds to the perpendicular magnetic field $B_{x}(x_\text{qd}(t))=\partial b_{\perp}\cdot x_\text{qd}(t)$ due to micro-magnets, responsible for Rabi driving. Here, $\sigma_x$ is the $x$ Pauli matrix acting on spin and $x_\text{qd} (t)$ is the position of the QD center at time $t$. The value of $\partial b_{\perp}$ for the SpinBus is typically chosen to be \SI[per-mode=symbol]{0.1}{\milli\tesla\per\nano\metre}. We consider at first the position of the electron varied as a simple sine wave given as
\begin{equation}
    x_\text{qd}(t) = x_0\sin{(\omega t + \phi)}.
    \label{eq:initial_pulse}
\end{equation}
with $x_0$, $\omega$ and $\phi$ denoting the amplitude, frequency and phase of the oscillation respectively. The SpinBus architecture provides dynamic control over the position $x_\text{qd}$ of the trapping potential, using the same finger gates responsible for shuttling the electron through the device, which directly translates into spatial control of the electron trapped inside the potential well. $H_{Zeeman}$ causes precession of the spin around the quantization axis, and $H_{Rabi}$ acts perpendicular to the quantization axis to rotate the spin in the preferred direction. The choice of the phase in Eq.~\ref{eq:initial_pulse} gives the usual $\sigma_x$ and $\sigma_y$ gates in the rotating frame. Rabi oscillations occur between the spin up and spin down states due to EDSR, with the frequency
\begin{equation}
    \omega_{\text{Rabi}} = g \mu_B \partial b_{\perp} x_0/\hbar,
    \label{eq:Rabi frequency}
\end{equation}
which is modified as $\Omega = \sqrt{\delta^2+\omega_{Rabi}^2}$ when there is a detuning $\delta$.
In general, the valley Hamiltonian can be expressed as:
\begin{equation}
H_{\text{valley}} = \Delta_{\text{real}}(x_\text{qd}(t))\tau _x + \Delta_{\text{imag}}(x_\text{qd}(t))\tau_y 
\end{equation}
where $\tau_x$ and $\tau_y$ are Pauli operators in the valley subspace, $\Delta_{\text{real}}(x_\text{qd}(t))$ and $\Delta_{\text{imag}}(x_\text{qd}(t))$ corresponds to the real and imaginary part of the complex position-dependent intervalley coupling. The magnitude of valley splitting is given by $E_{V}(x_\text{qd})=2\sqrt{\Delta_{\text{real}}^{2}(x_\text{qd}(t)) + \Delta_{\text{imag}}^{2}(x_\text{qd}(t))}$, and the valley phase by $\varphi_{V}=\arg(\Delta_{\text{real}}(x_\text{qd}(t)) + i \Delta_{\text{imag}}(x_\text{qd}(t)))$. The values of $\Delta_{\text{real}}(x_\text{qd}(t))$ and $\Delta_{\text{imag}}(x_\text{qd}(t))$ are generated according to two different models for the microscopic valley physics (cf. Fig.~\ref{fig:manipulation-zone} (b) and (c), respectively). The Ge-diffusion model considers the microscopic arrangement of Ge atoms inside the Si well, leading to larger or smaller valley splittings depending on the overlap of the electron wavefunction with Ge atoms. On the other hand, the step model describes the effect of interface miscuts arising from the growth process of the heterostructure. This model assumes the region between two miscuts to be smooth and to have a uniform intervalley coupling, but the presence of a miscut changes the effective ground and excited state of the valley before and after the miscut. An extensive treatment of both models can be found in \cite{alessandro}.

The last term in the Hamiltonian is a position dependent effective ZZ-interaction between the valley and spin DOF, given by
\begin{equation}
    H_{\text{spin-valley}} = -\kappa_z\left(\tilde{\tau}_{z}(x_\text{qd}) \otimes\sigma_{z}\right)
\end{equation}
where $\kappa_z$ is typically chosen to be on the order of \qtyrange[range-units=single,range-phrase=--]{1e-6}{5e-6}{\milli\electronvolt}, which is proportional to the relative $g$-factor variation $\delta g/g$ \cite{ruskovElectronFactor2018, rancicElectricDipole2016}. Here, $\tilde{\tau}_{z}(x_\text{qd})=\{\cos(\varphi_{V}(x_\text{qd}))\tau_x + \sin(\varphi_{V}(x_\text{qd}))\tau_y\}$ is the $z$ Pauli matrix acting on the ground and excited states of valley, and $\varphi_{V}(x_\text{qd})$ is the valley phase at a chosen starting position.

\subsection{Dynamics and Optimization}
\label{sec:Dynamics&Optimization}
The combined spin-valley system follows open system dynamics, with the valley DOF decaying in contact with the crystal environment and the spin DOF dephasing because of charge noise. The density matrix of the combined system, $\rho(t)$, evolves according to the master equation
\begin{multline}
    \label{eq:master}
    \frac{\mathrm{d}\rho}{\mathrm{d}t} (t) = -\frac{i}{\hbar} [H \left ( x_\text{qd}(t) \right ), \rho (t)] \\
        + \frac{1}{T_{1,v}} \mathcal{D} \left [ \tilde{\tau}_{-} \left ( x_\text{qd}(t) \right ) \right] (\rho (t))
        + \frac{1}{T_{2,s}} \mathcal{D} [ \sigma_z ] (\rho (t)),
\end{multline}
where we have the dissipative operator $\mathcal{D}[L](\rho) = L \rho L^\dagger - (\rho L L^\dagger + L L^\dagger \rho) / 2$, $T_{1,v}=~$\SI{100}{\nano\second} (corresponding to valley decay), and $T_{2,s}=$~\qtyrange[range-units=single, range-phrase=--]{20}{80}{\micro\second} (corresponding to spin dephasing). The jump operator used for the valley relaxation is given by \cite{alessandro}
\begin{equation}
    \tilde{\tau}_{-} (x_\text{qd}) = \begin{pmatrix}
        1 & e^{-i \varphi_{V}(x_\text{qd})} \\
        e^{i \varphi_{V} (x_\text{qd})} & -1
    \end{pmatrix},
\end{equation}
which is the lowering operator from the local valley excited state to the local valley ground state.

\subsubsection{GRAPE with JAX}

The trajectory of electron as shown in Eq. \ref{eq:initial_pulse} is discretized into piecewise-constant functions as in the usual approach of the well known GRAPE algorithm \cite{khanejaOptimalControl2005}. The time step is chosen so as to capture the relevant spin-valley dynamics, and is calculated to be around \SI{0.8}{\pico\second} from the average valley splitting along the length of the device. The discretized trajectory is used to perform state evolution and finally the average gate fidelity is calculated as explained in Sec. \ref{sec:Cost_function}. Performance improvements for state evolution and fidelity evaluation are achieved using the JAX library in python \cite{JAXComposable2018}. This also enables auto-differentiation of the cost function, which is essential for back-propagating the errors to perform optimization. The gradients found using auto-differentiation are used with the L-BFGS optimization method in SCIPY \cite{virtanenSciPyFundamental2020}, to minimize the average gate infidelity ($1-\bar{F}$). The optimization is done for a lesser number of controls (fixed to \num{10} controls per \unit{\nano\second}) than the time-discretized steps, and then interpolated while calculating the final gate fidelity \cite{motzoiOptimalControl2011}.

\subsubsection{Shaping function}
The initial control pulse is sinusoidal, as in Eq.~\ref{eq:initial_pulse}. To ensure that the pulse starts and ends at the same point, we shape the pulse using a Gaussian ramp envelope given as
\begin{equation}
G(t) = \begin{cases}
      f_{s} (t-t_r), & \text{if}~t\leq t_r \\
      1, & \text{if}~t\geq t_r \wedge t\leq T_g-t_r \\
      f_{s} (\tilde{t}+t_r), & \text{if}~t\leq T_g-t_r 
   \end{cases}
\end{equation}
where $f_{s} (t)=\alpha [\exp (-t^2/2\sigma^2) - \exp(-t_r^2/2\sigma^2)]$, with $\tilde{t}=t-t_f$, $\sigma=t_r/4$ and $\alpha=[1-\exp(-t_r^2/2\sigma^2)]^{-1}$. Here $t_r$ is the rise time, set here to \SI{1}{\nano\second}, and $T_g$ is the gate time. The shaped pulse to be optimized will be $\tilde{x}_\text{qd}(t)=G(t)\cdot x_\text{qd}(t)$.

\subsubsection{Cost function}
\label{sec:Cost_function}
The function to be maximized is the average gate fidelity of the quantum channel $\mathcal{E}$, evolving an initial spin state according to Eq.~\ref{eq:master} and subsequently tracing out the valley DOF, compared to a desired unitary $U$. We use 
\begin{equation}\label{eq:cost}
    \Bar{F}(\mathcal{E}, U)=\frac{dF_{\text{ent}}(\mathcal{E}, U)+1}{d+1},
\end{equation}
as defined in \cite{nielsenSimpleFormula2002}, with $d = 2$. Here, $F_{\text{ent}}(\mathcal{E}, U)$ is the entanglement fidelity and is denoted by $F_{\text{ent}}(\mathcal{E}, U)=\bra{\phi}(\mathds{1}\otimes\,\mathcal{U}^\dagger\circ\mathcal{E})(\phi)\ket{\phi}$, where $\ket{\phi}$ is a maximally entangled state of the spin with an ancilla qubit and $\mathcal{U}^{\dagger} (\rho) = U^\dagger \rho U$. In practice, the calculation of $F_{\text{ent}}(\mathcal{E})$ involves the propagation of four initial spin states that are orthogonal in the Bloch sphere. 

For our simulations, the target gate $U_G$ was chosen to be a rotation of $\pi$ about the y-axis, given by
\begin{equation}
    U_G = \exp\left(-i\pi\sigma_{y} / 2\right).
\end{equation}
As the simulated evolution in Eq.~\ref{eq:master} is in the lab frame, the spin precesses under the action of $H_{\text{Zeeman}}$ and $H_{\text{spin-valley}}$, in addition to the desired gate. In the ideal case of the electron remaining always in the valley ground state throughout the evolution, the frequency of precession is $\hbar\Omega_R = 2 \mu_B B_z + 2 \kappa_z$. We counter-rotate the final spin state to transform into a precession-free rotating frame via $U_{R} = \exp{\left(-i \Omega_R \sigma_{z}\right)}$. Finally in Eq.~\ref{eq:cost} we use $U = U_R U_G$ which ensures that the optimizer will favor the trajectory that keeps the electron in the valley ground state.

\begin{figure*}
    \centering
    \subfigure{\includegraphics[width=0.9\textwidth]{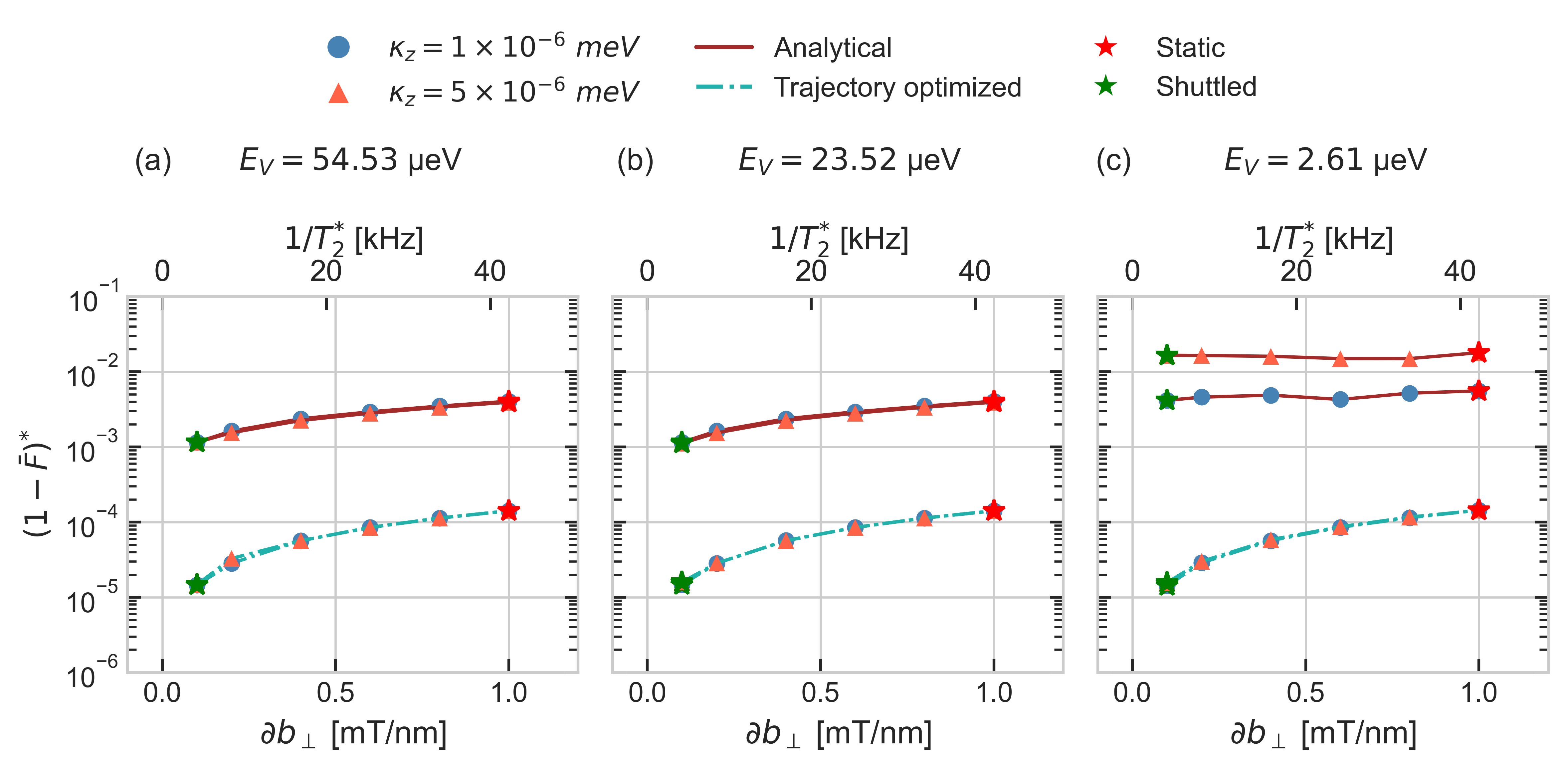}}
    \caption{\textbf{Effect of transverse magnetic gradient on spin dephasing and average gate fidelity for varying valley splittings.} 
    Ge-diffusion model is used to generate a prototypical device, and three points along the device are sampled. Pulses with varying $x_0$ (and corresponding analytical $T_g$) are used to obtain the average gate infidelity ($1-\Bar{F}$) at each point. The minimum of $1-\Bar{F}$ with respect to $x_0$ (indicated by *), for varying $\partial b_{\perp}$ and corresponding spin dephasing rates are plotted here. (a-b) Large valley splitting helps at retaining good fidelity as spin dephasing rate increases (red line). (c) LVSP combined with large spin-valley coupling leads to an order of magnitude worse average gate fidelity (red line with triangles). In all cases, the advantage of shuttling the qubits over operating them static can be seen by comparing their infidelities, as indicated by the green ($\partial b_{\perp}=$~\SI[per-mode=symbol]{0.1}{\milli\tesla\per\nano\metre}) and red ($\partial b_{\perp}=$~\SI[per-mode=symbol]{1}{\milli\tesla\per\nano\metre}) stars respectively. Irrespective of the competing physical mechanisms causing decoherence, optimal control always converges below \num{e-3}, with significant advantage from shuttling (teal dash-dot line).} 
    \label{fig:T2_compare}
\end{figure*}

\section{Results}\label{Sec:Results}

This section is divided into three parts. We first compare shuttling-based EDSR with static EDSR, by relating the influence of transverse magnetic field gradients on spin-dephasing and average gate fidelity. Next, we investigate the spin-valley interplay mechanism and the use of optimal control as a mean to perform high fidelity gates for driving pulses ranging in amplitude and duration. Finally, we show the differences in single-qubit gate fidelity for competing valley models. Device-specific and ensemble-averaged gate fidelities are analysed to understand the limitations imposed by spin-valley decoherence.

\begin{figure}\label{fig1}
    \subfigure{\includegraphics[width=0.5\textwidth]{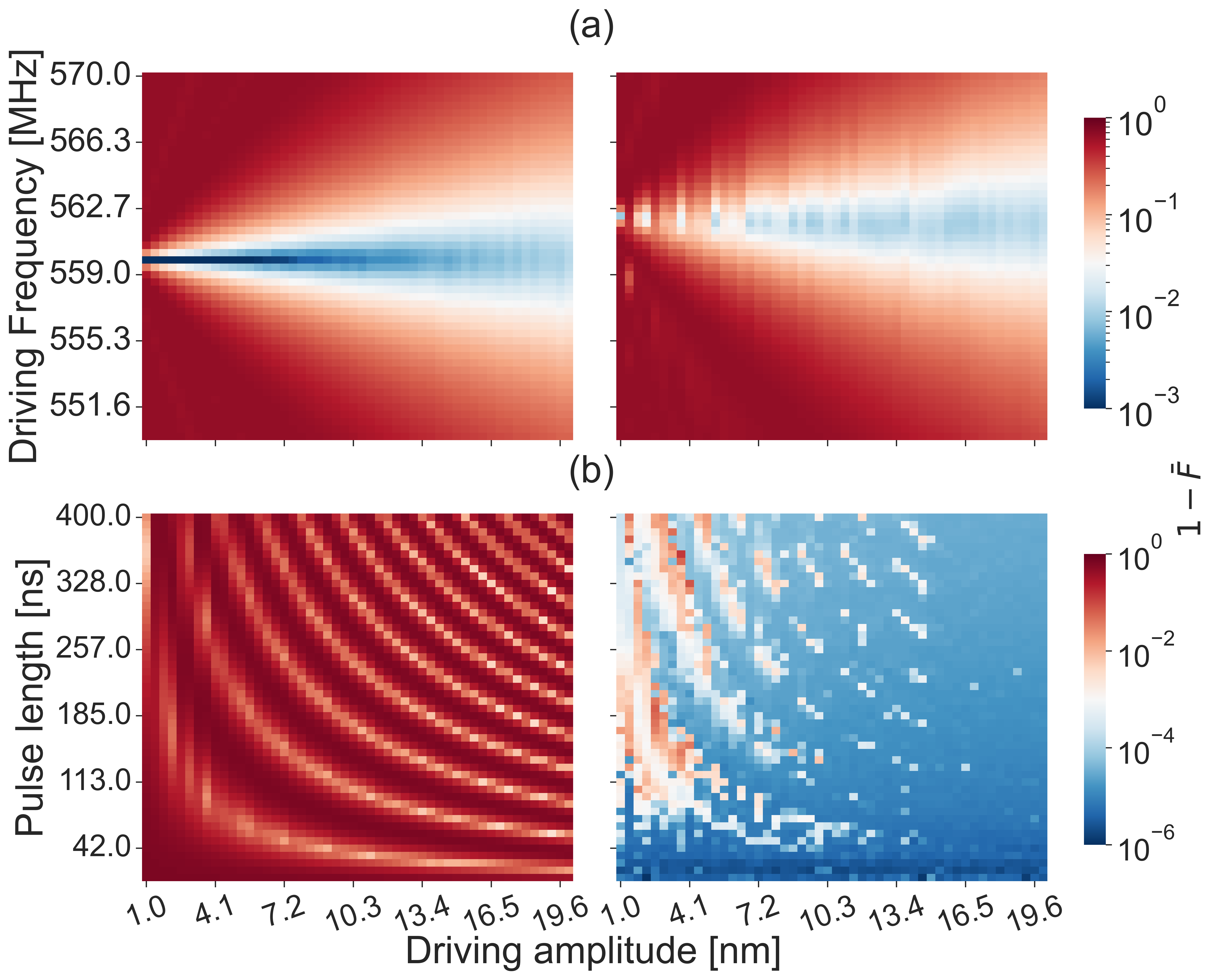}}
    \caption{\textbf{Trajectory parameters and optimization.} Amplitude of the driving pulse is swept along with (a) driving frequency and (b) pulse length to calculate the average gate infidelity, for a device simulated using Ge-diffusion model with spin-valley coupling $\kappa_z=$~\SI{5e-6}{\milli\electronvolt}. (a) left panel shows the case with no spin-valley coupling. When the driving frequency is closer to the Rabi frequency, higher fidelity is achieved. In the presence of spin-valley coupling the Rabi frequency is modified (visible from the upwards shift), and LZSM interference of valley excited states leads to fringes, as shown in the right figure of (a). The distinctive Chevron patterns are visible in (b), with no parameter combinations yielding infidelities below \num{e-3} (left panel). Pulses obtained through optimal control performs better at preserving the average gate fidelity, minimizing infidelities to well below \num{e-3} (right panel).}
    \label{fig:amp_sweep}
\end{figure}

\subsection{Static vs. shuttled quantum dots: Effect on spin dephasing}
\label{sec:Spin_dephasing}

Static quantum dots in a Si/SiGe heterostructure have achieved high fidelity single-qubit operations using micro-magnet enabled EDSR, with a $T_2^*=$~\SI{20}{\micro\second} \cite{yonedaQuantumdotSpin2018}, while operating in the regime mentioned in Sec.~\ref{sec:Device_architecture}. EDSR based on lower amplitudes limits achievable gate time, resulting in slower gates. In principle, larger $x_0$ combined with large $\partial b_{\perp}$ could lead to faster gates, but keeping small gradients reduces the effects of charge noise.

Charge noise arising from voltage fluctuations in confinement potentials is transferred to the spin as dephasing noise by the longitudinal component of the magnetic gradient, $\partial b_{\parallel}$. This reduces the $T_2^*$ of a device, which is derived in Ref.~\cite{kawakamiElectricalControl2014} as $T_2^* = 2\sqrt{\ln(2)}/(\pi \delta f_{\text{FWHM}})$, where $\delta f_{\text{FWHM}}=g \mu_B \partial b_{\parallel} \delta x_{\text{rms}}/h$, with $\delta x_{\text{rms}}$ denoting the root mean square displacement of the electron position $x$ caused by voltage fluctuations (see Ref.~\cite{yonedaQuantumdotSpin2018} supplementary). This yields
\begin{equation}
    T_2^* = \frac{2h\sqrt{\ln(2)}}{\pi g \mu_B \partial b_{\parallel} \delta x_{\text{rms}}}.
    \label{eq:spin-dephasing}
\end{equation}
Changing the geometry of the micro-magnet can maximize the ratio $Q=\partial b_{\perp}/\partial b_{\parallel}$, between transverse and longitudinal gradients, to improve $T_2^*$ of a device \cite{dumoulinstuyckLowDephasing2021}. Although design-dependent, maximum and minimum values of $Q$ are in general finite and in the range of one order of magnitude. It follows that the choice of transverse gradient used for EDSR will influence spin dephasing. Substituting $\partial b_\parallel = \partial b_\perp / Q$ in Eq.~\ref{eq:spin-dephasing} we obtain
\begin{equation}
    T_2^* = \frac{2hQ\sqrt{\ln(2)}}{\pi g \mu_B \partial b_{\perp} \delta x_{\text{rms}}}
    =\frac{4QT_{g}x_0\sqrt{\ln(2)}}{\pi \delta x_{\text{rms}}},
\end{equation}
directly relating the $T_2^*$ with $Q$. The amplitude $x_0$ and the analytical gate time $T_g$ computed as
\begin{equation}
    T_g = \frac{\pi}{\omega_{Rabi}} = \frac{\pi\hbar}{g\mu_B\partial b_\perp x_0}\,.
\end{equation}
In Ref.~\cite{yonedaQuantumdotSpin2018}, it is estimated that $\delta x_{\text{rms}}=$~\SI{4}{\pico\metre}, and the quantum dot is operated at $\partial b_{\perp}=$~\SI[per-mode=symbol]{1}{\milli\tesla\per\nano\metre} and $\partial b_{\parallel}=$~\SI[per-mode=symbol]{0.2}{\milli\tesla\per\nano\metre}. In order to study the effect of $\partial b_{\perp}$ on $T_2^*$, we relate these quantities by fixing $Q=5$. In the case of EDSR by shuttling, as one reaches an order of magnitude higher amplitudes, the transverse gradient can be $\partial b_{\perp}=$~\SI[per-mode=symbol]{0.1}{\milli\tesla\per\nano\metre} \cite{langrockBlueprintScalable2023} and therefore we assume $\partial b_{\parallel}=$~\SI[per-mode=symbol]{0.02}{\milli\tesla\per\nano\metre}.

We use the Ge-diffusion model to generate a valley landscape, and base our study on two different spin-valley couplings, specifically $\kappa_z=$~\SI{1e-6}{\milli\electronvolt}(typically expected value \cite{ruskovElectronFactor2018}) and $\kappa_z=$~\SI{5e-6}{\milli\electronvolt}. Three points are sampled from this valley landscape, corresponding to a valley splitting of \SI{54.53}{\micro\electronvolt}, \SI{23.52}{\micro\electronvolt} and \SI{2.61}{\micro\electronvolt} respectively. For each point, $\partial b_{\parallel}$ is varied from \qtyrange[per-mode=symbol,range-units=single, range-phrase=--]{0.02}{0.2}{\milli\tesla\per\nano\metre}, yielding the corresponding $\partial b_{\perp}$ and $T_2^*$ (see Eq.~\ref{eq:spin-dephasing}). It follows from Eq.~\ref{eq:Rabi frequency} that the Rabi frequency is directly proportional to $\partial b_{\perp}x_0$. For each value of $\partial b_{\perp}$, we vary the analytical $T_{g}$ from \qtyrange[range-units=single, range-phrase=--]{5}{360}{\nano\second} to generate electron trajectories with the corresponding analytical $x_0$ and calculate the average gate infidelity. From this, the minimum of the infidelity, $(1-\Bar{F})^*$, with respect to $x_0$ is calculated (Fig.~\ref{fig:T2_compare}, brown solid line). The same procedure is repeated for optimized electron trajectories (as explained in Sec.~\ref{sec:Dynamics&Optimization}) corresponding to each $T_{g}$ and $x_0$ (Fig.~\ref{fig:T2_compare}, teal dash-dot line).

The following observations are in order: First, smaller magnetic gradients results in smaller spin dephasing rates for relatively larger values of valley splittings as shown by the brown solid lines in Fig. \ref{fig:T2_compare}(a)-(b). The advantage of smaller gradients wanes at a significantly lower valley splitting, denoting the importance of spin-valley physics at larger amplitudes, as shown in Fig. \ref{fig:T2_compare}(c). The effect of larger spin-valley couplings only shows a noticeable difference in the case of a small valley splitting. Second, optimal control of electron trajectories ensures an average gate infidelity below \num{e-3}, irrespective of the magnitude of the valley splitting or spin-valley coupling for all magnetic gradients.
\begin{figure*}
    \centering
    \subfigure{\includegraphics[width=0.94\textwidth]{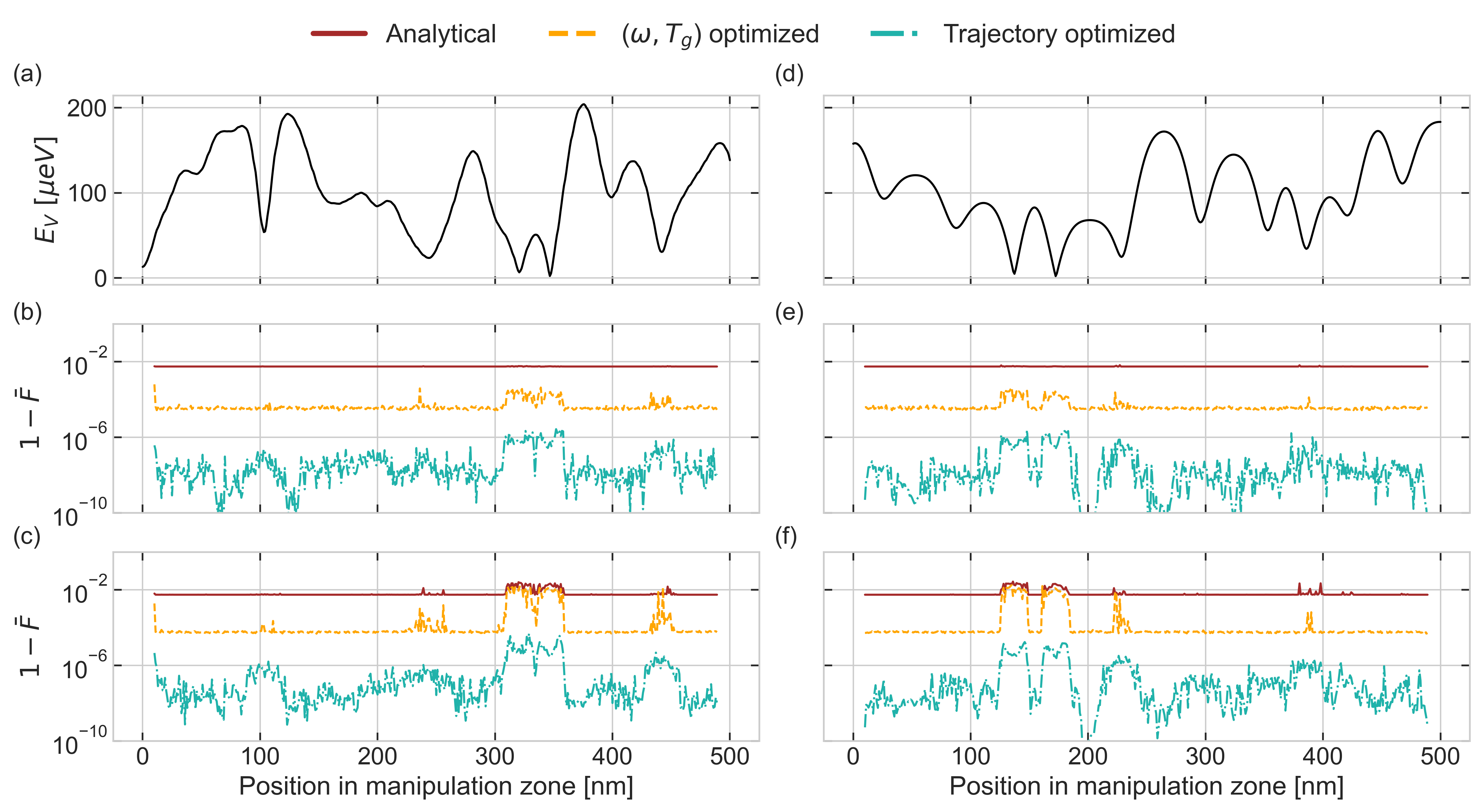}}
    \caption{\textbf{Dependence of average gate fidelity on micro-magnet position in the manipulation zone.} For a device simulated using the Ge-diffusion model, the position of micro-magnet is varied over a chosen length. (a) Valley splitting dependence on position. (b) Analytical (brown line), ($\omega, T_g$)-optimized (orange dashed line) and trajectory optimized (teal dash-dot line) infidelity as a function of position, for $\kappa_z=10^{-6}$ meV. As the valley splitting encountered decreases, the infidelity increases and becomes worse when it encounters a LVSP. Optimal control improves the average gate fidelity even at hard-to-navigate LVSPs. (c) For $\kappa_z=5\times 10^{-6}$ meV, this effect is enhanced for the analytical and ($\omega, T_g$)-optimized lines, while optimal control succeeds at retaining high fidelity. (d), (e) and (f) shows the same for the step model of valley splitting.} 
    \label{fig:Device-compare}
\end{figure*}

\begin{figure*}
  \centering
  \subfigure{\includegraphics[width=0.9\textwidth]{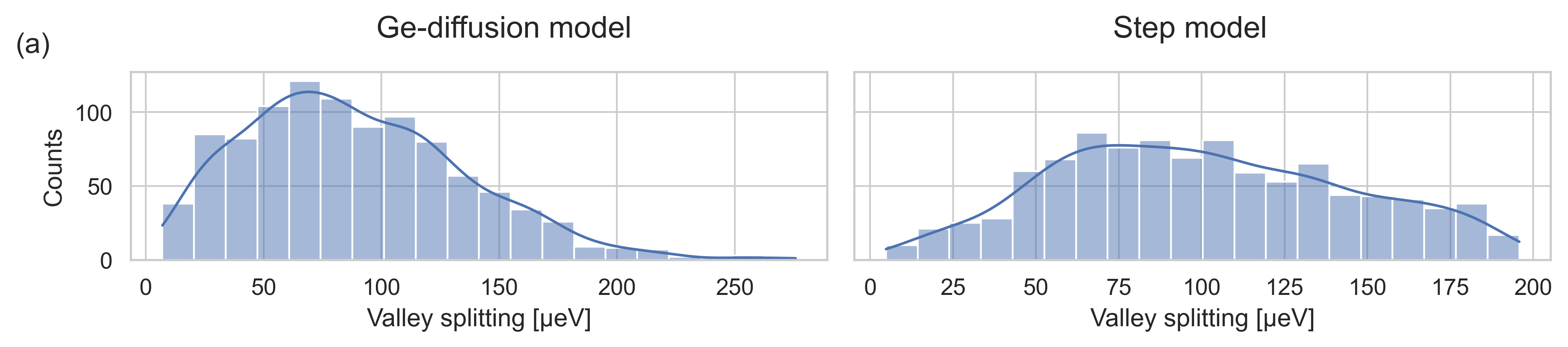}\label{fig:Ge_samples_a}}\vspace{0pt}
  \subfigure{\includegraphics[width=\textwidth]{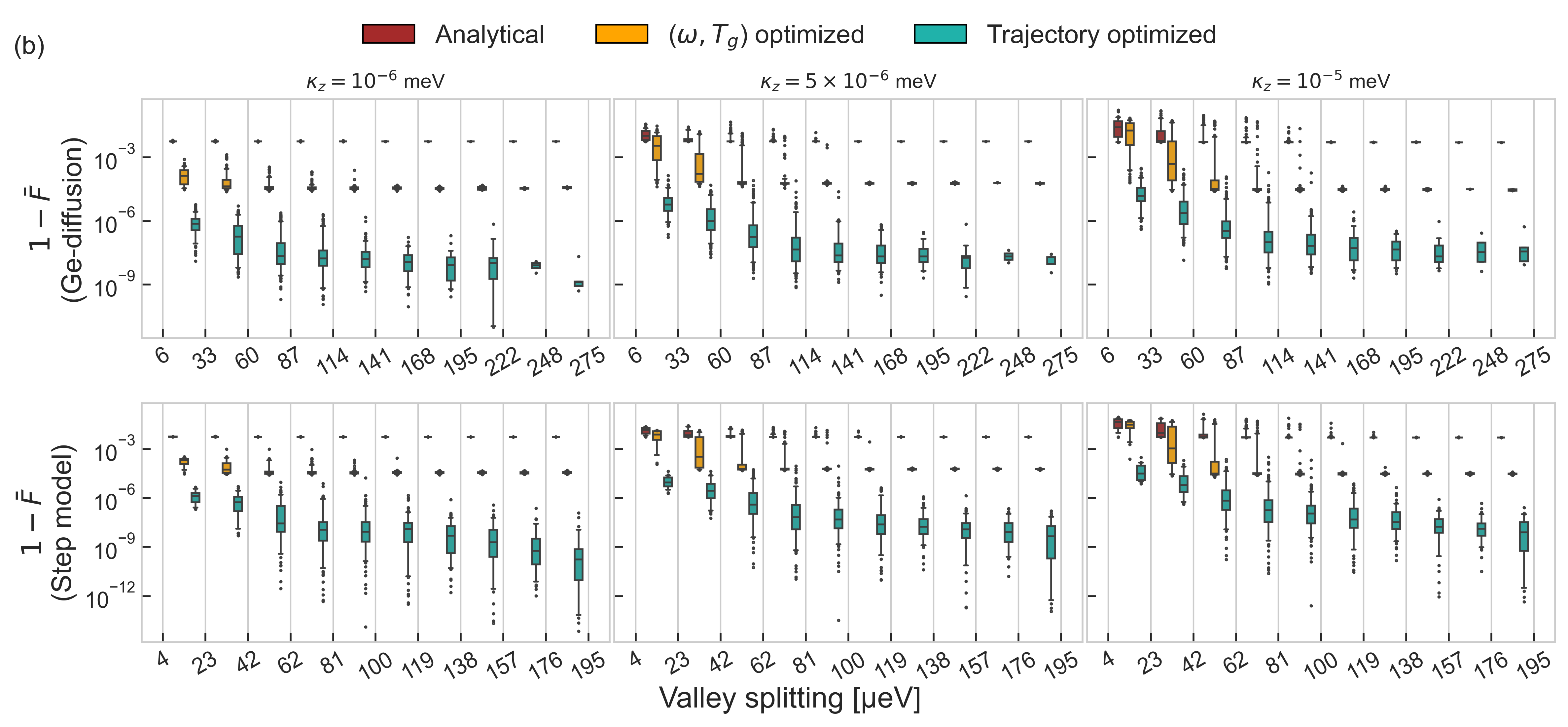}\label{fig:Ge_samples_b}}
  \caption{\textbf{Optimal control improves average gate fidelity over \num{1000} sampled devices.} Ge-diffusion model and step model are used to generate 1000 different shuttling devices of \SI{200}{\nano\metre} length, with varying valley splitting distributions. (a) Valley splitting distribution calculated at the center of each device (at \SI{100}{\nano\metre}). (b) Bar plots between average gate infidelity and valley splitting with mean marked for Ge-diffusion model (top row) and step model (bottom row). Increasing values of spin-valley coupling leads to poor average gate fidelity for LVSPs, as shown by the analytical infidelity bars (in brown). ($\omega,T_g$) optimization can be employed to improve fidelity for larger valley splittings, but is limited by LVSPs (orange bars). Pulse engineering helps in improving the average gate fidelity by at least three orders of magnitude even after encountering LVSPs (teal bars).}
  \label{fig:Ge_samples}
\end{figure*}

\subsection{Dependence of average gate fidelity on pulse parameters}

Electron trajectories can be engineered for a variety of amplitudes ($x_0$), frequencies ($\omega$) and pulse lengths ($T_g$). To study how the system responds to these different parameters, a Ge-diffusion model based device is used to create a well defined valley landscape, with $\kappa_z=$~\SI{5e-6}{\milli\electronvolt}, and $E_{V}=103\pm49$\,\unit{\micro\electronvolt}. A LVSP where the valley splitting is $\sim$\SI{7}{\micro\electronvolt} is chosen along the length of the device for Rabi pulsing to demonstrate the advantage of optimal control. The micro-magnet configuration chosen in Ref.~\cite{kunneSpinBusArchitecture2023} is predicted to limit the maximum amplitude of oscillations to \SI{20}{\nano\metre} to perform EDSR.

Sweeping the values of $\omega$ for $x_0$ up to \SI{20}{\nano\metre}, we calculate the average gate infidelity in the absence and presence of spin-valley coupling, as shown in Fig. \ref{fig:amp_sweep}(a). The left panel of the figure shows the case where $\kappa_z=$~\SI{0}{\milli\electronvolt}, with optimal fidelities obtained for $\omega\sim560$ \unit{\mega\hertz}, corresponding to an external field of $B_z=$~\SI{20}{\milli\tesla}. Increasing spin-valley coupling to $\kappa_z=$~\SI{5e-6}{\milli\electronvolt} (right panel) shows a shift in resonant $\omega$ (to  $\sim562.3$ \unit{\mega\hertz}) leading to optimal fidelities. The well known Landau-Zener-Stückelberg-Majorana (LZSM) interference \cite{shevchenkoLandauZener2010}, which causes phase accumulation between excited- and ground state when passing an anti-crossing point in energy periodically, happens in the valley DOF. A strong spin-valley coupling results in fringes along the frequency band. The fringes are predominant for amplitudes $<$ \SI{10}{\nano\metre}, suggesting that valley excitations are limited at larger amplitudes.

A sweep of $x_0$ and $T_g$ is also performed, and the entire parameter grid is optimized as mentioned in Sec.~\ref{sec:Dynamics&Optimization} for \num{300} iterations. The left panel of Fig. \ref{fig:amp_sweep}(b) shows the parameter grid before optimization, with no possible parameter combination yielding desired fidelities. Fringes due to LZSM interference are visible at lower amplitudes, as noted above. The right panel of Fig. \ref{fig:amp_sweep}(b) shows the same parameter grid after trajectory optimization, leading to \qty{88.68}{\percent} of the parameter grid yielding infidelity below \num{e-3}, highlighting the potential improvement from using optimal control. For all amplitudes, pulse lengths $<$\SI{42}{\nano\second} resulted in high fidelity gates. Meanwhile, driving amplitudes $>$\SI{14}{\nano\metre} yielded high fidelity gates for all pulse lengths. This suggests that even though there is a wide range of parameters where the optimal control performs better, specific ranges of amplitudes and pulse lengths are more desirable when optimizing the average gate infidelity.

\subsection{Valley-model-agnostic optimal control}

Valley models are differentiated by the choice of sampling the real and imaginary part of the complex position dependent valley splitting, as mentioned in Sec.~\ref{Sec:Background}. Ge-diffusion and step models of valley splitting are plausible models that can be used to simulate realistic conveyor-mode shuttling based quantum dot devices. This section is divided into two parts. The first part deals with the positioning of the micro-magnet along the length of the device, and the next part deals with sampling valley profiles of \num{1000} different devices. In both parts, we show that very low average gate infidelities can be obtained, irrespective of the valley model used. In particular, optimal control always ensures trajectories that results in high fidelity single-qubit gates for the SpinBus architecture. 

\subsubsection{Micro-magnet position dependence}
\label{sec:MM_position}
A prototypical device is simulated using both valley splitting models, with $\kappa_z=$~\SI{1e-6}{\milli\electronvolt} and $\kappa_z=$~\SI{5e-6}{\milli\electronvolt}. In a realistic scenario, Fig.~\ref{fig:Device-compare}(a) and (d) shows the valley landscape for Ge-diffusion and step model respectively. At each point along the valley landscape, a sinusoidal electron trajectory with amplitude $x_0=$~\SI{10}{\nano\metre} and corresponding analytical $T_{g}$ at driving frequency $\omega\sim$\,\SI{562.3}{\mega\hertz} is used to calculate average gate infidelity. Empirically, tuning $\omega$ to account for shifts due to spin-valley dynamics, and tuning $T_g$ to get rid of any over-rotation or under rotation of the spin while performing the desired gate, could improve the average gate fidelity in experiments. In a real experiment, calibration of $\omega$ and $T_g$ could be done independently to find the sweet-spot for attaining large gate fidelities. This process is numerically simulated using Bayesian optimization based on Gaussian processes \cite{garnett_bayesoptbook_2023}, implemented in the python library \emph{scikit-optimize} \cite{headScikitoptimizeScikitoptimize2021}. This method is faster than the brute-force parameter sweeps for calibration, since we start the optimization in a well defined initial interval around $\omega_{\text{Rabi}}$ and analytical $T_g$. Conservatively, fidelities in a real experiment might fall between the analytical and Bayesian optimized distributions due to limitations in precision of evaluating fidelities, for respective valley splittings. Furthermore, the initial trajectory based on analytical $T_{g}$ is optimized as discussed in Sec.~\ref{sec:Dynamics&Optimization}. 

The results of optimization for Ge-diffusion and step model corresponding to $\kappa_z=$~\SI{1e-6}{\milli\electronvolt} is shown in Fig.~\ref{fig:Device-compare}(b) and (e), and corresponding to $\kappa_z=$~\SI{5e-6}{\milli\electronvolt} are shown in Fig.~\ref{fig:Device-compare}(c) and (f), with red solid line denoting the analytical infidelity, orange dotted line denoting ($\omega, T_g$)-optimized infidelity and teal dash-dot line denoting trajectory-optimized infidelity.

It is observed that the gate infidelities are valley landscape dependent, which makes it really important to choose an appropriate operating point for placing the micro-magnet to increase the probability of performing a high fidelity single-qubit operation in SpinBus. As noted in Sec.~\ref{sec:Spin_dephasing}, larger values of valley splitting inherently help in achieving high fidelities, even in the presence of an unfavourable spin-valley coupling. This is evident from the flat profile for the ($\omega, T_g$)-optimized infidelity at larger valley splitting values, irrespective of the spin-valley coupling strength. It can also be noted that the gradient of valley splitting with respect to the position works in conjunction with the valley splitting strength to decrease average gate fidelity (for instance, between \qtyrange[range-units=single, range-phrase=--]{200}{300}{\nano\metre} in Fig.~\ref{fig:Device-compare}(c)), due to the LZSM interference. Spin decoherence due to LVSPs are more prevalent at larger spin-valley couplings. The difference in the valley models chosen does not seem to change the analytical and post-optimization fidelity characteristics. 

Calibrating $\omega$ and $T_g$ might suffice to provide competing fidelities for a vast majority of points along the valley landscape as shown in Fig.~\ref{fig:Device-compare}b)-(f). However, trajectory optimization always lets the average gate fidelity converge below \num{e-4} despite encountering competing physical effects discussed above.

\subsubsection{Device sampling}

The valley splitting models are used to generate 1000 devices of same length, with varying valley landscape. We assume that the center of the micro-magnet is placed exactly at the center of the \SI{200}{\nano\metre} long device (at \SI{100}{\nano\metre}).

\emph{ Ge-diffusion model}: As mentioned in Sec.~\ref{Sec:Background}, a valley profile is generated by relative positioning of Ge atoms in the Si well. The presence of Ge atoms alters the intervalley coupling, causing a spatially varying difference in energy. In order to sample devices simulated using the Ge-diffusion model, a seed is chosen to randomly place Ge atoms along the Si well. Fig \ref{fig:Ge_samples}(a) (left panel) shows the valley splitting profile for the sampled devices. The average valley splitting at the center taken over the \num{1000} devices is $\sim88\pm46$ \unit{\micro\electronvolt}, while the average splitting over the full length of the \num{1000} devices is $\sim88\pm17$ \unit{\micro\electronvolt}. The distribution of valley splittings is skewed towards lower splittings, with \qty{53.9}{\percent} of the samples below a splitting of $\sim$88 µeV. Valley splittings below \SI{33}{\micro\electronvolt} are encountered with \qty{12.3}{\percent} probability.  

\emph{ Step model}: The relative placement of miscuts along the device is tuned using a chosen seed so as to generate different valley profiles. The corresponding valley splitting profile is shown in Fig. \ref{fig:Ge_samples}(a) (right panel). The average valley splitting at the center is $\sim 100\pm44$ \unit{\micro\electronvolt}, while the average splitting over thefull length of the \num{1000} devices is $\sim 92\pm15$ \unit{\micro\electronvolt}. The valley splittings are skewed more towards the larger values, with \qty{56.6}{\percent} of the samples above a splitting of $\sim$\SI{88}{\micro\electronvolt}. Valley splittings below \SI{33}{\micro\electronvolt} are encountered with \qty{5.6}{\percent} probability, relatively less than the Ge-diffusion model. This could be attributed to lower intervalley coupling being concentrated around the step edges.

For both the models, we employ the same methodology as discussed in the previous sub-section (Sec.~\ref{sec:MM_position}) to calculate analytical, ($\omega, T_g$)-optimized and trajectory optimized gate infidelities, with the electron always oscillating about the center of the device for each sample. Fig.~\ref{fig:Ge_samples}(b) shows the results for Ge-diffusion (top row) and step model (bottom row) after optimization. We observe that the analytical Rabi pulses result in poor gate infidelities as the valley splitting decreases below \SI{60}{\micro\electronvolt}, progressively getting worse for lower valley splittings (cf. Fig. \ref{fig:Ge_samples}(b), brown bars). Calibrating $\omega$ and $T_g$ can improve the fidelity for a vast majority of samples (orange bars), but is limited by LVSPs and relatively larger spin-valley coupling values. It can be easily seen that optimized electron trajectories performs at least three orders of magnitude better than the analytical Rabi pulses for all values of valley splittings. 

It is evident that an important factor that plays a role in limiting the average gate fidelity other than the valley splitting strength is the spin-valley coupling. Gradually increasing the spin-valley coupling shifts the attainable infidelities upwards, implying more spin decoherence, in general. For $\kappa_z=$~\SI{1e-6}{\milli\electronvolt}, none of the samples using analytical pulses converged below an average gate infidelity of $10^{-3}$, while 779 (Ge-diffusion) and 899 (step model) out of \num{1000} samples using ($\omega, T_g$)-optimized pulses converged below an average gate infidelity of \num{e-3}. Trajectory shaping using optimal control for the same leads to an average gate infidelity below \num{e-3} for all samples irrespective of the underlying valley model or spin-valley coupling, posing a significant advantage as opposed to the usual Rabi pulses. In conclusion, detrimental effects of encountering LVSPs can be effectively mitigated by optimal control.

\section{Discussion}
\label{sec:Discussion}

Conveyor-mode shuttling based architectures can be engineered with better spin-dephasing times than the static quantum dot counterparts due to its ability to perform larger amplitudes of electron position oscillation. We have shown that for realistic spin decoherence rates, an order of magnitude improvement over standard EDSR may be attainable, which makes the SpinBus architecture a potential candidate for fault-tolerant quantum computation \cite{fowlerSurfaceCodes2012}. 

Oscillating the electron at larger amplitudes increases the chance of valley excitations and positionally-dependent frequency offsets, thus hindering spin coherence and high fidelity gates. The existence of LVSPs can severely limit fidelity, to the extent to make these manipulation zones unusable. In this work, we have shown that such a drastic measure is unneeded, and applying optimal control still permits error rates below \num{e-3}, enabling full use of the SpinBus architecture. While we have focused here on model-based control, most of the conclusions here should still hold in the case of in-situ closed-loop control. 

Meanwhile, we have shown that the valley DOF, with sufficiently large splitting ($>$\SI{60}{\micro\electronvolt}) does not hamper the spin significantly unless the spin-valley coupling is pessimistically large. However, proper calibration to take into account position-dependent frequency shifts is needed. The single-qubit gate fidelities after calibration can be used to probe the unknown valley landscape along a device, to map out potential LVSPs. This can be used to shift the choice of the initial operating position of the electron to yield high fidelity gates without the use of optimal control. Full optimal control enables high fidelity gates even if there is a chance to encounter such spin-valley couplings in a realistic device. We have modeled both the Ge-diffusion and step model scenarios in detail, and shown that in both cases, even under pessimistic conditions, control theory can greatly enhance device operability. It is in principle also possible to engineer electron oscillation trajectories that are inherently robust to the differences in valley landscapes, as mapping out the valley landscape for large scale devices might not be practical.

Moreover, the uncertainties in measuring valley splittings in an experimental scenario could be mitigated using closed loop control techniques, when the trajectories could be modified to yield optimal gate fidelities real-time. Since the average number of iterations to yield high fidelity gates for the open loop scheme discussed in this work is around \num{100}, extending it to a closed loop control scheme might increase the iterations by roughly \num{10} times \cite{rossignoloQuOCSQuantum2023}, which still is a break even trade-off to get high fidelity gates, especially given fast duty cycles for these devices.

It is also important to note that there might be residual nuclear spins from $^{29}$Si even after isotopical purification, which could cause further dephasing for the spin. Using basic driving frequency and gate time optimization in such a scenario might not yield high fidelity gates. However these once again can be straightforwardly improved, e.g. using dynamical decoupling \cite{violaDynamicalSuppression1998, yonedaQuantumdotSpin2018}, and by using the optimal control schemes presented in this work.

\section{Conclusion}\label{Sec:Conclusion}

A quantum bus architecture paired with micro-magnets is promising for large scale universal quantum computation. We have quantitatively shown that order-of-magnitude improvement over standard EDSR is possible for shuttling based EDSR. Nonetheless, the choice of the material stack for the device (Si/SiGe in this work) presents additional complexity in the Hamiltonian and decay channels that cause decoherence of the information encoded in the spins of the electron. Higher gate fidelities are limited by the so called valley degree of freedom arising from the conduction band minima of Si, which is further limited by the charge noise mediated through the micro-magnets used for single-qubit manipulation. 

We show that straightforward control theoretic routines can improve the gate fidelities by at least three orders of magnitude compared to analytical formulas. These results are tested for two competing models for the description of valley splittings, namely Ge diffusion and the step model. Even for valley splittings below \SI{15}{\micro\electronvolt}, which would otherwise be essentially unusable for computation, the optimized pulses would result in \qty{99.9}{\percent} gate fidelity, thus improving controllability in tough-to-navigate LVSPs that could be present along the length of the device.

\begin{acknowledgments}

This work was supported by the Helmholtz Validation Fund project “Qruise” (HVF-00096), and by the Germany’s Excellence Strategy – Cluster of Excellence Matter and Light for Quantum Computing (ML4Q) EXC 2004/1 – 390534.
\end{acknowledgments}

\bibliography{main.bib}

\end{document}